\documentclass[12pt,epsbox]{article}
\usepackage{color}
\usepackage[dvips]{graphicx,psfrag}
\usepackage{amsmath,amssymb}
\usepackage{bm}
\baselineskip=\normalbaselineskip
\renewcommand{\baselinestretch}{1.4}
\newcommand{\resection}[1]
 {\setcounter{equation}{0}\section{\large{#1}}}

\renewcommand{\thefootnote}{\fnsymbol{footnote}}
\newcommand{\bel}[1]{\begin{equation}\label{#1}}
\newcommand{\bal}[1]{\begin{eqnarray}\label{#1}}

\newcommand{\be}{\begin{equation}}
\newcommand{\ee}{\end{equation}}
\newcommand{\ba}{\begin{eqnarray}}
\newcommand{\ea}{\end{eqnarray}}
\newcommand{\nn}{\nonumber \\}

\newcommand{\ran}{\rangle}
\newcommand{\lan}{\langle}

\newcommand{\vev}[1]{\left\langle\,{#1}\,\right\rangle}
\newcommand{\tr}{{\rm tr}}

\newcommand{\hphi}{\hat{\phi}}

\newcommand{\eq}[1]{(\ref{#1})}

\newcommand{\del}{\partial}

\begin{document}
\setcounter{page}{0}
\begin{flushright}
\parbox{40mm}{%
KUNS-1852 \\
{\tt hep-th/0307029} \\
July 2003}

\end{flushright}

\vfill

\begin{center}
{\large{\bf 
Effects of space-time noncommutativity \\
on the angular power spectrum of the CMB
}}
\end{center}

\vfill

\begin{center}
{\sc Masafumi Fukuma}\footnote%
{E-mail: {\tt fukuma@gauge.scphys.kyoto-u.ac.jp}},  
{\sc Yuji Kono}\footnote%
{E-mail: {\tt kono@gauge.scphys.kyoto-u.ac.jp}} and 
{\sc Akitsugu Miwa}\footnote%
{E-mail: {\tt akitsugu@gauge.scphys.kyoto-u.ac.jp}}  \\[2em]
{\sl Department of Physics, Kyoto University, Kyoto 606-8502, Japan} \\

\end{center}

\vfill
\renewcommand{\thefootnote}{\arabic{footnote}}
\setcounter{footnote}{0}
\addtocounter{page}{1}

\begin{center}
{\bf abstract}
\end{center}

\begin{quote}

We investigate an inflationary model of the universe 
based on the assumption 
that space-time is noncommutative in the very early universe. 
We analyze the effects of space-time noncommutativity 
on the quantum fluctuations of an inflaton field 
and investigate their contributions to the cosmic microwave background (CMB). 
We show that the angular power spectrum $l(l+1)C_l$ 
generically has a sharp damping for lower $l$ 
if we assume that the last scattering surface
is traced back to fuzzy spheres at the times when 
large-scale modes cross the Hubble horizon.
\end{quote}
\vfill
\renewcommand{\baselinestretch}{1.4}
\newpage
\resection{Introduction}
Recently the WMAP research group presented their first year result 
on the CMB anisotropy 
\cite{Bennett:2003bz}\cite{Spergel:2003cb}\cite{Peiris:2003ff}. 
They determined many of the relevant cosmological parameters 
with great precision from the angular power spectrum $C_l$, 
which is defined by 
\begin{align}
 \left\lan \frac{\delta T}{T}(\eta_0,\Omega_1)\,
  \frac{\delta T}{T}(\eta_0,\Omega_2)\right\ran 
  =\sum_l \frac{2l+1}{4\pi}\, C_l \,P_l(\cos\theta_{12}). 
\end{align}
Here $\eta$ is the conformal time, 
and $(\delta T/T)(\eta_0,\Omega)$ is the CMB temperature fluctuation 
observed at the present time ($\eta\!=\!\eta_0$) 
in the angular direction $\Omega=(\theta,\varphi)$.  
$l$ is the azimuthal quantum number, and $\theta_{12}$ is the
angle between the directions $\Omega_1$ and $\Omega_2$. 
The expectation value $\vev{~}$ represents the sample average 
over the data taken from various parts on the celestial sphere.

The data of WMAP and COBE \cite{cobe} show that for small $l$ ($\lesssim 50$), 
$l(l+1)C_l$ take values almost constant in $l$,  
which can be naturally explained 
if the curvature perturbation has an almost scale-invariant power spectrum. 
This has been regarded as a strong evidence 
for inflationary models of the universe \cite{Guth:1981}\cite{Sato1981}.%
\footnote{
See refs.\ \cite{Gratton:2003pe}\cite{Khoury:2003vb}\cite{Fukuma:2003kv} 
for recent attempts to derive 
a scale-invariant power spectrum based on models 
other than inflationary models. 
}   
In fact, the leading slow-roll approximation exactly yields  
a scale-invariant power spectrum,
and moreover, small deviations from the scale-invariant power spectrum 
can be generically accommodated 
by adjusting the potential of inflaton(s).

However, $l(l+1)C_l$ starts to deviate from the constant value 
as $l$ becomes much smaller ($l< 10$). 
This behavior is difficult to be explained with standard inflationary models.%
\footnote{
See refs.\ \cite{Yokoyama:1998rw}\cite{Kawasaki:2003dd} 
for attempts to derive this behavior 
solely from inflationary models.
} 
Conventional explanation of this deviation 
is based on the so-called cosmic variance. 
That is, for smaller $l$, 
any theoretical calculations inevitably get to loose their predictive power. 
In fact, what one can predict theoretically is only the mean value of
some large ensemble,  
but for small angular momentum modes we can get only a few experimental data  
because a smaller $l$ corresponds to a larger angular scale.

If one takes the deviation seriously, however,
then it may be regarded as 
a sign of the necessity to change our understanding 
of the fundamental dynamics  
in the very early universe. 
The main purpose of the present article is to show 
that this sharp damping could be understood 
as a generic property that always holds 
when the space-time noncommutativity is incorporated 
into the dynamics of an inflaton field around the string scale.%
\footnote{%
Other effects of Planck-scale physics on the CMB anisotropy 
are also expected to be found, 
such as a running spectral index or a deviation from
the Gaussian perturbation. 
See refs.\ \cite{Chu:2000ww}
for attempts in this direction. 
}

\begin{figure}[htbp]
\begin{center}
\resizebox{!}{60mm}{
\input{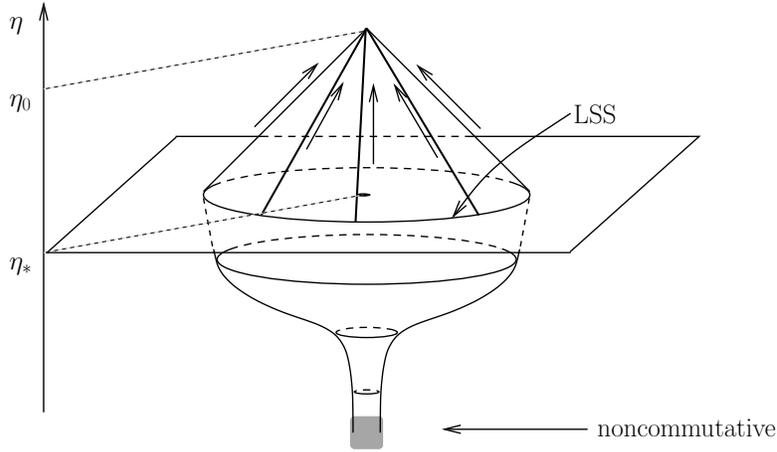}
}
\caption{\footnotesize{
An observer at the present time $\eta = \eta_0$ sees photons from the
LSS as the CMB. The LSS is traced back to a noncommutative sphere in
the very early universe. 
}}
\label{LSS}
\end{center}
\end{figure}%
The point of our discussion is the following (see Fig.\ \ref{LSS}). 
The observed CMB mainly consists of photons coming from 
the last-scattering surface (LSS)  
which is a two-sphere on the time slice (at $\eta=\eta_\ast$) 
when the recombination takes place. 
This sphere can be traced back to spheres at earlier times ($\eta<\eta_\ast$)
such that they have the same comoving coordinates with those 
of the sphere at $\eta_\ast$ (the LSS). 
This family of spheres makes an orbit in four-dimensional space-time, 
and we assume that spheres become fuzzy or noncommutative 
as they come back to the very early universe. 
On the other hand, the angular power spectrum $C_l$ 
is proportional to the corresponding power spectrum of the
gravitational potential on the LSS 
as a result of the Sachs-Wolfe effect \cite{Sachs}. 
This gravitational potential  in turn is related 
to the amplitude that the corresponding inflaton mode 
takes when it crosses the Hubble horizon. 
As we see later, $l$ plays almost the same role 
with that of a comoving wave vector $k$ 
in evaluating the power spectrum ($k\sim l$), so that
a mode with smaller $l$ crosses the Hubble horizon at an earlier time. 
Therefore, under our assumption of the noncommutativity 
in the very early universe, 
modes with smaller $l$ see more noncommutativity of  space-time, 
leading to a modification of the angular power spectrum for smaller $l$. 

In the present article, we make a rough estimation 
of the angular power spectrum,
taking into account only the noncommutativity of the angular 
coordinates $\Omega = (\theta, \phi)$ since these should yield the most
relevant effects on the deviations of $l(l+1)C_l$ from the standard value.%
\footnote
{See refs.\ \cite{Brandenberger:2002nq}\cite{Huang:2003zp}\cite{q} 
for discussions that take into account the noncommutativity 
for $\eta$ and/or $r$.
} 
In fact, in the cosmological perturbation theory, 
one can consider the time evolution of each mode separately. 
Furthermore, the angular power spectrum of the CMB anisotropy 
can be related to the fluctuations of gravitational potential 
on the last scattering surface, 
and thus is sensitive to the fluctuations only in the angular directions. 
Since the noncommutativity to a given direction 
is expected to give its major effects to the fluctuations 
in the corresponding direction, 
the introduction of noncommutativity to the other directions 
$(\eta, r)$ 
will not give a drastic change to the angular power spectrum.%
\footnote{
Recently Tsujikawa, Maartens and Brandenberger \cite{Tsujikawa:2003gh} 
and Huang and Li \cite{Huang:2003hw} 
have analyzed the noncommutative inflation 
introducing the noncommutativity only to $(\eta,r)$, 
and shown that the effect is not strong enough 
to give a sharp damping to $l(l+1)C_l$ at large angular scales. 
}
We show that when the noncommutativity is introduced to angular directions,  
$l(l+1)C_l$ with small $l$ certainly has a substantial 
deviation from the constant value. 
A more detailed analysis is possible once extra settings are 
incorporated correctly, 
and will be reported in the forthcoming paper \cite{fkm2}.

This paper is organized as follows. 
In section 2 we give an inflationary model of the universe,
assuming that the very early universe has a space-time noncommutativity 
only for the angular coordinates.  
In section 3 we calculate the CMB anisotropy based on our model, 
and show that $l(l+1)C_l$ has a sharp damping for small $l$.
Section 4 is devoted to conclusion and outlook. 

\resection{Model}

The flat Friedmann-Robertson-Walker (FRW) universe 
in the absence of noncommutativity is given by the metric 
\begin{align}
 ds^2 =g_{\mu\nu}\,dx^\mu\,dx^\nu =a(\eta)^2\,
  \bigl(-d\eta^2+d\vec{x}\,{}^2\bigr)
  = a(\eta)^2\,\bigl(-d\eta^2 + dr^2 + r^2 d\Omega^2\bigr).  
\end{align}
In the present article, we exclusively use the conformal time $\eta$ 
to represent the time coordinate.
During the inflationary era, 
the size of the universe, $a(\eta)$, is given by
\begin{align}
 a(\eta)= -\frac{1}{H\eta} 
\end{align}
with the constant Hubble parameter $H$. 
The action of an inflaton field $\Phi(x)=\Phi(\eta,\vec{x})
=\Phi(\eta,r,\Omega)$ 
with the potential $V(\Phi)$  is given by 
\begin{align}
  S[\Phi(x)]&=\int d\eta \,dr \,d\Omega\, \sqrt{-g}
   \Bigl[\,-\frac{1}{2}\,\nabla_\mu \Phi(x) \nabla^\mu \Phi(x)
    -  V(\Phi)\,\Bigr].
\end{align}
In order to describe the Gaussian fluctuations of the inflaton field 
around the classical value $\bar{\phi}(\eta)$, 
we set $\Phi(x)=\bar{\phi}(\eta)+\phi(x)$ 
and expand the above action around $\bar{\phi}(\eta)$ 
to the quadratic order:%
\footnote{
In the following, we neglect the contribution from the metric perturbation 
since it can be shown to be sufficiently small during the inflationary era.
} 
\begin{align}
 &{\bf S}\bigl[\bar\phi(\eta)+\phi(x)\bigr] 
  \simeq{\bf S}\bigl[\bar\phi(\eta)\bigr]+ S\bigl[\phi(x)\bigr] \label{ac}\\
 &S\bigl[\phi(x)\bigr]=\int d\eta\, dr\, d\Omega\, a^2(\eta)\, r^2\,
  \Bigl[\,-\frac{1}{2}\,\eta^{\mu\nu}\,
  \partial_\mu\phi(x)\,\partial_\nu \phi(x) 
  - \frac{1}{2}\,a^2(\eta)\,\phi^2 \,V''(\bar\phi)\,\Bigr],
 \label{flac}
\end{align}
where $\eta_{\mu\nu}=\textrm{diag}[-1,+1,+1,+1]$.

Now we introduce a noncommutativity into space-time. 
As stated in Introduction, 
we introduce the noncommutativity only to the angular coordinates 
$\Omega=(\theta,\varphi)$,  
so that the three-dimensional space is the product 
of the radial direction $r$ and the fuzzy sphere.

Assuming that an observer is at the origin ($r=0$), 
we consider a sphere of radius $r$ around the observer.  
Since $r$ is the radius of the comoving coordinates, 
the physical radius $\rho$ of the sphere changes under 
the evolution of the universe as 
\begin{align}
 \rho(\eta,r)=a(\eta)\,r. 
\end{align}
This implies that if we introduce a noncommutativity on spheres 
by allowing at most one bit of degrees of freedom to
reside in the unit physical Planck area $l_s^2$, 
then the number of degrees of freedom 
(or the dimension of the Hilbert space on which field operators 
can act) 
also changes as the universe develops.

The above statement can be made into a precise form in the following way. 
We first introduce a fuzzy sphere 
\cite{hppe1982}\cite{Hoppe:vv}\cite{deWit:1988ig} 
such that it is represented by the noncommutative comoving coordinates 
$\hat{x}_i$ ($i=1,2,3$) 
with the relation 
\begin{align}
 \left[\hat{x}_i , \hat{x}_j\right]&= i\,\theta \,\varepsilon_{i j k}\,
  \hat{x}_k \,\,\,\,,\,\,\,\,
  \sum_{i=1}^3 \left(\hat{x}_i\right)^2 = r^2 \;.
\end{align}
We assume that the Hilbert space on which $\hat{x}_i$ acts 
has dimension $(N+1)$. 
Then, $\hat{x}_i$ can be represented by the generators $\hat{L}_i$ 
in the $(N+1)$-dimensional (i.e., spin-$(N/2)$) representation 
of the $su(2)$ algebra as 
\begin{align}
 \hat{L}_i = \theta^{-1} \hat{x}_i,
\end{align}
where
\begin{align}
 \left[\hat{L}_i,\hat{L}_j\right] &= i\,\varepsilon_{i j k}\,\hat{L}_k  
  \,\,\,\,,\,\,\,\,
  \sum_{i=1}^3 \bigl(\hat{L}_i\bigr)^2 
  = \frac{N}{2}\left( \frac{N}{2}+1 \right)\;.
\end{align}
Here we should note that $\theta=2r/\sqrt{N(N+2)}$ should not be regarded 
as the fundamental noncommutative scale,  
since $\theta$ simply represents the noncommutativity 
of the comoving coordinates. 
A correct interpretation is given as follows \cite{Madore}. 
Since three coordinates $\hat x_i$ cannot be diagonalized simultaneously, 
we can best characterize the position of a point on the fuzzy sphere 
only with an eigenvalue of one coordinate, say $\hat{x}_3$. 
Thus in the $(N+1)$-dimensional representation, 
the sphere consists of $(N+1)$  fundamental regions 
(because $\hat{L}_3 = \theta^{-1}\hat{x}_3$ has $(N+1)$ eigenvalues).
Since each fundamental region should have the area of noncommutative scale, 
$l_s^2$, we have the following relation for a sphere of 
{\em physical} radius $\rho$: 
\begin{align}
 \frac{4\pi\rho^2}{N+1} = l_s^2. 
 \label{nonc}
\end{align}
In our model of the FRW universe, $\rho$ is a function of $\eta$ and $r$ 
as $\rho =a(\eta)r$, so that eq.\ (\ref{nonc}) defines $N$ as a function of
$(\eta,r)$ as 
\begin{align}
 N(\eta,r)=\frac{4\pi a^2(\eta) r^2}{l_s^2}-1.
 \label{noncer}
\end{align}

In order to analyze a scalar field theory on this fuzzy sphere, we
must translate usual functions on a smooth sphere into operators on this
fuzzy sphere. 
This can be carried out as follows.
We first define truncated spherical harmonics $\widehat{Y}_{lm}$:
\be
 \widehat{Y}_{lm}=\frac{a^l}{\rho^{\,l}}\sum_{i_k}
  f^{lm}_{i_1 i_2 ... i_l}\,\hat{x}_{i_1}\,\hat{x}_{i_2} \cdots \hat{x}_{i_l},
\ee
where the traceless  symmetric functions $f^{lm}_{i_1 i_2 ... i_l}$ 
are the coefficients appearing in the usual spherical harmonics: 
$Y_{lm}(\Omega) 
= (a^l/\rho^{\,l})\sum_{i_k}f^{lm}_{i_1 i_2  ... i_l}\,
x_{i_1}x_{i_2}\!\cdots x_{i_l}$. 
Note that for the spherical harmonics on the fuzzy sphere, 
$l$ is limited to $N$.%
\footnote{
The number of independent coefficients $f^{lm}_{i_1i_2...i_l}$ 
$(0\leq l\leq N)$ is given by 
$\sum_{l=0}^{N}\bigl({}_{l+2}\textrm{C}_2 - {}_l\textrm{C}_2\bigr) =
\sum_{l=0}^{N}(2l+1)=   (N+1)^2$. From this we can see that with
$f^{lm}_{i_1i_2...i_l}$ $(0 \leq l \leq  N)$ $\widehat{Y}_{lm}$ form 
a complete basis of $(N+1) \times (N+1)$ hermitian matrices.
}
By the definition of $f^{lm}_{i_1i_2...i_l}$, $\widehat{Y}_{lm}$
satisfies the following equations:
\begin{align}
\big[ \hat{L}_3 , \widehat{Y}_{lm} \big]& = m\widehat{Y}_{lm}
\,\,\,\,,\,\,\,\,
\sum_{i=1}^3\big[ \hat{L}_i ,[ \hat{L}_i , \widehat{Y}_{lm}]] = l(l+1)
\widehat{Y}_{lm}.
\label{yl}
\end{align}
Any function on a smooth sphere
can be expanded with spherical harmonics $Y_{lm}(\Omega)$:
\be
\phi(\Omega) = \sum_{l=0}^{\infty}\sum_{m=-l}^{l} \phi_{lm}Y_{lm}(\Omega).
\ee
Then using these coefficients $\phi_{lm}$ and truncated spherical
harmonics, we define the operator $\hphi$  on the fuzzy sphere  
that corresponds to $\phi(\Omega)$ as
\begin{align}
\hat{\phi} \equiv \sum_{l=0}^N \sum_{m=-l}^{l}
 \phi_{lm} \widehat{Y}_{lm}.
\end{align}
Taking into account the remaining commutative coordinates $(\eta, r)$, 
every function $\phi(\eta,r,\Omega)$ in a smooth space-time 
becomes a matrix-valued two-dimensional field
\begin{align}
\hat{\phi}(\eta,r) = \sum_{l=0}^{[N(\eta,r)]} \sum_{m=-l}^{l}
 \phi_{lm}(\eta,r)\,\widehat{Y}_{lm}.
\label{trfn}
\end{align}
We here used the Gauss symbol $[N(\eta,r)]$ because (\ref{noncer})
defines $N$ as a real number.
The important point is that the mode expansion is bounded from above
by $N$ that is a function of $(\eta,r)$.
That is, the two-dimensional field $\hat{\phi}(\eta,r)$ 
can have the mode expansion only up to the maximal value 
\begin{align}
 l_{max}=N(\eta,r)=\frac{4\pi a^2(\eta)\,r^2}{l_s^2}-1
\end{align}
at a point with the space-time coordinates $(\eta,r)$. 
Since $a(\eta)=-1/H\eta$ during the inflationary era, 
the above equation implies that 
for fixed $r$ the mode of $l$ newly appears at the moment 
$\eta_l(r)\equiv -(r/l_sH)\sqrt{4\pi/(l+1)}$.

We now rewrite the scalar field action (\ref{flac}) 
on this noncommutative space-time:
\begin{align}
S\bigl[\hat\phi\bigr] = \frac{1}{2}
\int& d\eta\, dr\, a^2(\eta)\, r^2
\frac{1}{N+1}\,\tr\biggl[\,\bigl(\partial_\eta {\hphi}\bigr)^2 
 - \bigl(\partial_r \hphi\bigr)^2
 + \frac{1}{r^2}\,\bigl[\hat L_i,\hphi\bigr]^2
 -a^2(\eta)\,{\hphi}{}^2\,V''(\bar{\phi})\,\biggr]
\nn
=\frac{1}{2}
 \int& d\eta \,dr\, a^2(\eta)\,r^2\, \sum_{l=0}^{\left[N(\eta,r)\right]}
  \sum_{m=-l}^{l} \nn
 &\times\left[\,\bigl|\partial_\eta{\phi}_{lm}\bigr|^2 -
  \bigl|\partial_r\phi_{lm}\bigr|^2 -
  \frac{l(l+1)}{r^2}\,|\phi_{lm}|^2
  -a^2(\eta)\,V''(\bar\phi)\,|\phi_{lm}|^2\right].
\end{align}
Here we replaced $\int d\Omega/4\pi$ with $1/(N+1)\,\tr$, and
$i\,\varepsilon _{ijk}\,x_j\,\partial_k$ with $\bigl[\hat L_i,\,\bigr]$.
In the second line, we used the normalization
$
\tr \,\widehat{Y}^\dagger_{lm} \widehat{Y}_{l'm'}
  =(N+1)\,\delta_{l l'}\,\delta_{m  m'} 
$
and eqs.\ (\ref{yl}), neglecting possible contributions from the boundary.
Rewriting the summation as
\begin{align}
\int d\eta \,dr \sum_{l=0}^{[N(\eta,r)]}\sum_{m=-l}^{l} = \int d\eta \,dr
\sum_{l=0}^{\infty}\sum_{m=-l}^{l} \theta\bigl(N(\eta,r)\!-l\bigr),
\end{align}
we obtain the final form of the action:
\begin{align}
S\bigl[\hphi\bigr] 
 =\frac{1}{2}\,&\int
 d\eta\,dr\,a^2(\eta)\,r^2 \,\sum_{l=0}^{\infty}\sum_{m=-l}^{l}\,
 \theta\bigl(N(\eta,r)\!-l\bigr) \nn
&\times\left[\,\bigl|\partial_\eta{\phi}_{lm}\bigr|^2 -
  \bigl|\partial_r\phi_{lm}\bigr|^2 -
  \frac{l(l+1)}{r^2}\,|\phi_{lm}|^2
  -a^2(\eta)\,V''(\bar\phi)\,|\phi_{lm}|^2\right].
\label{fac}
\end{align}

\resection{Analysis of fluctuations}

In this section, we calculate the angular power spectrum $C_l$ for small $l$.
This is related to the large-scale gravitational potential 
on the LSS through the Sachs-Wolfe effect.
By studying the Einstein equation, it turns out that the gravitational 
potential on the LSS 
is determined by the inflaton fluctuations \cite{Liddle}.
We consider only the Gaussian fluctuations, 
so the inflaton fluctuations can be calculated 
by its two-point function. 
The equation of motion (EOM) for the mode $\phi_{lm}(\eta, r)$ 
of an inflaton in a noncommutative background is obtained from the action 
\eqref{fac} as
\begin{align}
\left( \del_\eta^2 - \del_r^2 + \frac{l(l+1)}{r^2} - \frac{2}{\eta^2}\right) 
\bigl( a(\eta)\,r\, \phi_{lm}(\eta,r) \bigr) =0. \label{3.EOM}
\end{align}
Here we used the equality $ \del_\eta^{\,2} a/a = 2/\eta^2$.
We also adopted the slow-roll approximation 
and neglected the term $V''(\bar{\phi})$.

The EOM should be solved with an appropriate boundary condition 
at the boundary $\eta=-(r/l_s H)\,\sqrt{4\pi/(l+1)}$ 
in order to calculate the two-point function of $\phi_{lm}(\eta, r)$ 
correctly. 
However, 
in order to understand the qualitative behavior of the angular 
power spectrum for lower $l$, 
one may simplify the problem with the following assumptions: 
\begin{itemize}
\item Because the inflaton field in the superhorizon is frozen 
to the value it takes when crossing the Hubble horizon,
those modes that are allowed only in the superhorizon should take the
vanishing value also in the superhorizon.
\item Those modes that are allowed in the subhorizon should have 
the same behavior with that in the commutative case.
\end{itemize}

In the commutative limit, the EOM \eqref{3.EOM} can be solved as 
\begin{align}
 \phi_{lm}(\eta, r) &= \int_0^\infty dk\ H\eta^2 \sqrt{\frac{k^3}{\pi}}\, 
 j_l(kr) 
 \left[ b_{lm}(k)\,h_1^{(1)}(-k\eta)  
 + (-1)^m b^\dag_{l\,-m}(k)\,h_1^{(2)}(-k\eta)\right]\nn
 &\equiv \int_0^\infty dk \,
  \sqrt{\frac{2}{\pi}}\, k\, j_l(kr) \,\phi_{lm}(\eta, k).
\label{3.sol}
\end{align}
Here $h_{1}^{(1)}$ and $h_{1}^{(2)}$ are the spherical Hankel functions 
of order 1, 
and the coefficients $b_{lm}(k)$ satisfy the commutation relations
\begin{align}
 \bigl[\,b_{lm}(k),\,b_{l'm'}^\dagger(k')\,\bigr]
  =\delta(k-k')\,\delta_{ll'}\,\delta_{mm'},
\end{align}
which are equivalent to the canonical commutation relations 
for the field $\phi_{lm}(\eta,r)$: 
\begin{align}
 &\bigl[\,\phi_{lm}(\eta,r),\,\pi_{l'm'}(\eta,r')\,\bigr]
  =i\,\delta(r-r')\,\delta_{ll'}\,\delta_{mm'},\\
 &\pi_{lm}(\eta,r)\equiv a^2(\eta)\,r^2\,\partial_\eta\phi^*_{lm}(\eta,r).
\end{align}
\begin{figure}[htbp]
\begin{center}
\resizebox{!}{60mm}{
 \input{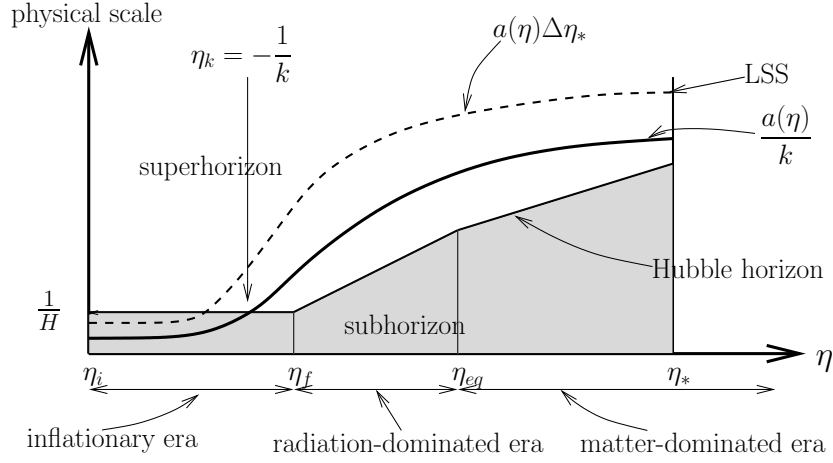}
}
\caption{\footnotesize{
During the inflationary era, 
the Hubble radius $1/H$ remains constant. 
The physical wave length $a(\eta)/k$ of a mode in the subhorizon 
crosses the Hubble horizon at $\eta=\eta_k$ 
and enters the superhorizon (unshaded in the figure). 
}}
\label{horscale}
\end{center}
\end{figure}%

\begin{figure}[htbp]
\begin{center}
\resizebox{!}{60mm}{
\input{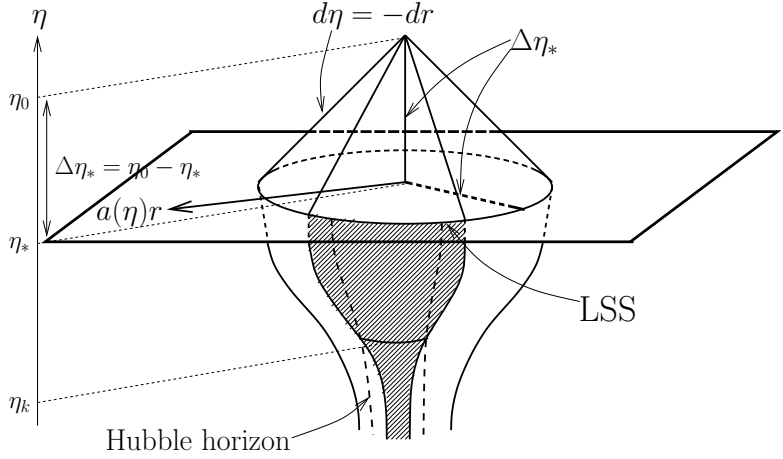}
}
\caption{\footnotesize{
The comoving distance between the observer and a point on the LSS 
is given by the difference of their conformal times, $\Delta\eta_*$.  
The development of the physical wave length of mode $k$ is depicted
(shaded region). 
A fluctuation of the Hubble scale at $\eta=\eta_k$ 
is transformed into the gravitational potential on the LSS.
}}
\label{horizon}
\end{center}
\end{figure}

We here consider a mode that has a wave number $k$ and an angular momentum $l$.
This mode crosses the Hubble horizon at $\eta = \eta_k\equiv -1/k$, 
where the physical wave length $a(\eta)/k$ equals the Hubble distance $1/H$ 
(see Fig.\ \ref{horscale}). 
Because of our assumptions above, 
the mode $\phi_{lm}(\eta,k)$ in the expansion (\ref{3.sol}) can 
have quantum fluctuations in the subhorizon 
only when the condition 
\begin{align}
 l\leq N(\eta_k, r)
\end{align}
is satisfied. 
Since we are interested in the fluctuations of the CMB, 
we should evaluate $N(\eta_k, r)$ at $r$ which corresponds to the LSS. 
As can be seen from Fig.\ \ref{horizon}, 
a photon travels from a point on the LSS to the observer 
along the line determined by $d\eta = -dr$, 
and thus the comoving distance between the observer ($r=0$) 
and the point on the LSS is given by 
the difference of their conformal times,  
$r=\eta_0-\eta_\ast\equiv \Delta \eta_*$ 
(see Fig.\ \ref{horizon}).
Therefore when a mode with wave number $k$ crosses the Hubble horizon, 
the angular-momentum cutoff is given by
\begin{align}
 N( \eta_k, \Delta \eta_*) 
  &= \frac{4\pi k^2 \bigl(\Delta \eta_*\bigr)^2 }{l_s^2H^2}-1.
\label{3.con}
\end{align}
Thus, for the noncommutative case, 
we should discard those modes in eq.\ \eqref{3.sol} that do 
not satisfy the condition 
$l\!\leq\! N(\eta_k,\Delta\eta_*)$.
This can be done by replacing  $b_{lm}(k)$ in eq.\ \eq{3.sol} with 
$b_{lm}(k)\,\theta\bigl(N(\eta_k, \Delta \eta_*)-l\bigr)$, 
and we obtain
\begin{align}
 \phi_{lm}(\eta, \Delta \eta_*) 
 &= \int_{0}^\infty dk\, 
 \sqrt{\frac{2}{\pi}}\,
 k\,j_l(k\Delta\eta_*)\,\phi_{lm}(\eta, k) \,
 \theta\bigl(N(\eta_k,\Delta\eta_*)-l\bigr)\nn
 &= \int_{k_*(l)}^\infty dk\, 
 \sqrt{\frac{2}{\pi}}\,
 k\,j_l(k\Delta\eta_*)\,\phi_{lm}(\eta, k) 
\end{align}
with
\begin{align}
  k_*(l) \equiv \frac{l_s H}{\Delta \eta_*} \sqrt{\frac{l+1}{4\pi}}.
\end{align}
{}From this expression we see that 
the noncommutativity during the inflationary era is translated 
into the infrared cutoff $k_*(l)$ of the $k$ integration.
After entering the superhorizon, 
the mode $\phi_{lm}(\eta,k)$ becomes ``classical" 
with the amplitude fixed to the value 
it takes when crossing the Hubble horizon. 
By setting $\eta=\eta_k$ in the two-point function, 
we thus obtain the power spectrum in the superhorizon,
which is almost constant in time during the inflationary era:  
\begin{align}
 \vev{\phi^*_{lm}(\eta_k,k)\,\phi_{l'm'}(\eta_{k'},k') }
  =\frac{2\pi^2}{k^3}\,P_\phi(k)\,\delta(k-k')\,
  \theta\bigl(k-k_\ast(l)\bigr)\,\delta_{ll'}\,\delta_{mm'}.
 \label{2pt_phi}
\end{align}
Here $P_\phi(k)$ is the power spectrum for the commutative case, 
which is given by 
\begin{align}
 P_\phi(k)=\frac{H^2}{4\pi^2}
 \label{2pt_phi2}
\end{align} 
in the leading slow-roll approximation.

Before calculating the angular power spectrum, 
we here make a comment on the mechanism 
through which a short-distance cutoff 
can affect the large-scale behavior of the CMB anisotropy. 
In our subsequent paper \cite{fkm2} we have investigated 
the possible ways of introducing cutoff into inflationary models 
such that it exhibits a sharp damping at large angular scales. 
There we have shown that the damping occurs as a result 
of the competition between two moments: 
one is the moment when a mode crosses the Hubble horizon 
and becomes a classical fluctuation, 
and the other is the moment when the mode is released 
from the constraint of cutoff. 
The angular power spectrum has a sharp damping 
at large angular scales when the first moment is prior 
to the second one only for larger-scale modes.  
In fact, the CMB angular power spectrum is related 
to the classical values of quantum fluctuations of inflaton, 
which will be largely suppressed when the above situation is realized,  
because then the large-scale modes must become classical 
before the modes start their quantum fluctuations. 
We also have shown there that this situation is realized when 
the noncommutativity is introduced to the angular directions.

We now calculate the angular power spectrum $C_l$. 
This can be carried out simply by following the usual prescription 
for obtaining $C_l$ from the two-point function of inflaton \cite{Liddle}. 
We first parametrize the metric under scalar perturbations 
in the longitudinal gauge 
\cite{Bardeen:kt}\cite{Kodama:bj}\cite{Brandenberger:1992}: 
\begin{align}
 ds^2 = a^2(\eta)\Bigl[ -\bigl(1 +2\Psi(\eta,\vec{x})\bigr) d\eta^2 +  
 \bigl(1 + 2\Phi(\eta,\vec{x})\bigr)\,\delta_{ij}\,dx^i dx^j \Bigr] 
\end{align}
with $\Psi$ and $\Phi$ being gravitational potentials. 
We assume that the cosmological perturbation 
\cite{Bardeen:kt}\cite{Kodama:bj}\cite{Brandenberger:1992}\cite{Liddle}  
is still applicable since the space-time noncommutativity disappears rapidly, 
so that the relation $\Phi = -\Psi$ holds 
when the anisotropic stress-tensor vanishes. 
If we further assume that the perturbations are adiabatic, 
then the combination ${\cal R}(\vec{x})= \Psi(\eta,\vec{x})
-(\partial_\eta a/a)\,v(\eta,\vec{x})$ 
gives a constant of motion on the superhorizon scale 
and is called the curvature perturbation. 
Here the velocity field $v(\eta,\vec{x})$ is defined  
through the $(\eta,i)$-component of the energy-momentum tensor as 
$T_{\eta i}=-a^2\bigl(\bar{\rho}+\bar{p}\bigr)\partial_i v$ 
with the unperturbed energy density and pressure, $\bar{\rho}(\eta)$ 
and $\bar{p}(\eta)$.  
We have ${\cal R}(\vec{x})=-(\partial_\eta a/a)\,v(\eta,\vec{x})
=H \phi(\eta,\vec{x})/\dot{\bar{\phi}}(\eta)$ 
during the inflationary era, 
and ${\cal R}(\vec{x})=(5/3)\Psi(\vec{x})$ during the matter-dominated era. 
Thus, by expanding $\Psi(\eta,\vec{x})=\Psi(\eta,r,\Omega)$  
with respect to the spherical harmonics 
(and further to the spherical Bessel functions) as%
\begin{align}
 \Psi(\eta,r,\Omega) = \sum_{l=0}^\infty\sum_m
   \Psi_{lm}(\eta, r) \,Y_{lm}(\Omega) 
 = \sum_{l=0}^\infty \sum_m
 \int_0^\infty dk\,\Psi_{lm}(\eta, k) \sqrt{\frac{2}{\pi}} \,
 k\,j_l(kr)\,Y_{lm}(\Omega),
\end{align}
the gravitational potential on the LSS, $\Psi_{lm}(\eta_\ast,k)$, 
is expressed as \cite{Liddle}
\begin{align}
 \Psi_{lm}(\eta_\ast,k)=\frac{3}{5}\,
  \frac{H}{\dot{\bar{\phi}}(\eta_k)}\,\phi_{lm}(\eta_k,k). 
\end{align}
The power spectrum $P_\Psi(k)$ of the gravitational potential, 
\begin{align}
 \vev{\Psi^\ast_{lm}(\eta_*,k)\,\Psi_{l'm'}(\eta_*,k') }
  =\frac{2\pi^2}{k^3}\,P_\Psi(k)\,\delta(k-k')\,
  \theta\bigl(k-k_\ast(l)\bigr)\,\delta_{ll'}\,\delta_{mm'},
 \label{2pt_Psi}
\end{align}
is then expressed as 
\begin{align}
 P_\Psi(k) &= \left( \frac{3H}{5 \dot{\bar{\phi}}(\eta_k)}\right)^2
  P_{\phi}(k)\nn 
 &\!\!\!\!\!\left(= \left( \frac{3H^2}{10 \pi \dot{\bar{\phi}}(\eta_k)}
  \right)^2
 \mbox{in the leading slow-roll approximation}\right).
 \label{2pt_Psi2}
\end{align}
{}Furthermore, through the Sachs-Wolfe effect \cite{Sachs}\cite{Liddle}, 
the temperature fluctuations in the CMB, 
\begin{align}
 \frac{\delta T(\eta, \Omega)}{T} &= \sum_{l=0}^\infty \sum_m
  a_{lm}(\eta)\,Y_{lm}(\Omega),
\end{align}
are related to 
the gravitational potential:
\begin{align}
 a_{lm}(\eta_0) &= \frac{1}{3}\,\Psi_{lm}(\eta_*, \Delta\eta_*)\nn
 &=\frac{1}{3}\,\int_0^\infty dk\,\Psi_{lm}(\eta_\ast,k)\,
  \sqrt{\frac{2}{\pi}}\,k\,j_l(k\Delta\eta_\ast)\nn
 &=\frac{1}{3}\,\int_{k_*(l)}^\infty dk\,\Psi_{lm}(\eta_\ast,k)\,
  \sqrt{\frac{2}{\pi}}\,k\,j_l(k\Delta\eta_\ast).
 \label{a_lm}
\end{align}

\begin{figure}[htbp]
\begin{center}
\rotatebox{270}{
\resizebox{!}{100mm}{
 \includegraphics{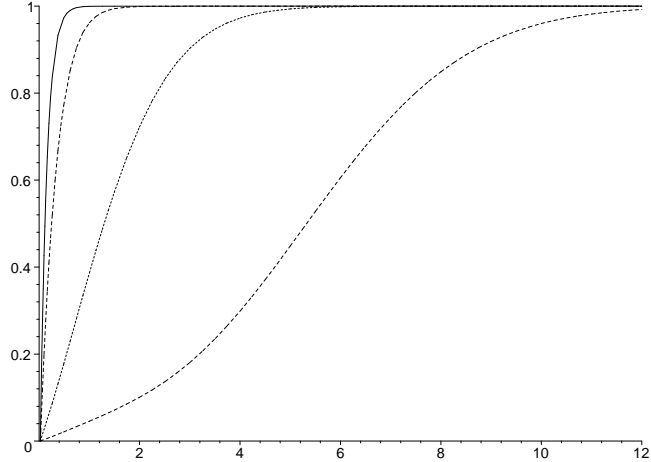}
}
}
\caption{\footnotesize{
Results of the calculation of the ratio $C_l/ C_l^{(0)}=1-\beta_l$. 
The vertical axis is the ratio $C_l/C_l^{(0)}$, 
and the horizontal axis is angular momentum $l$.
We set $n=0.95$ and change $l_s H$ from $0.1$ to $10$ (curves in the
figure have $l_sH = 0.1,\, 1,\, 5$ and $10$ from top to bottom).
There is a damping at small $l$, 
and the ratio approaches $1$ for large $l$.
}}
\label{N=0.95}
\end{center}
\end{figure}

We here recall that the spectral index $n(k)$ is defined as
\begin{align}
n(k)= \frac{d \log P_\Psi (k)}{d \log k} + 1.
\end{align}
In the superhorizon, $P_\Psi(k)$ depends on $k$ 
through $\dot{\bar{\phi}}(\eta_k)$ and also through the potential term 
which we neglected in solving the EOM, 
and thus, in the leading slow-roll approximation 
we have $n=1$, i.e., the power-spectrum $P_\Psi(k)$ is scale-invariant. 
However, when noncommutativity is taken into account, 
an IR cutoff is introduced into the $k$ integration. 
The angular power spectrum $C_l$ is calculated from eqs.\ 
\eq{2pt_Psi}, \eq{2pt_Psi2} and \eq{a_lm} as
\begin{align}
\vev{ a^*_{lm}(\eta_0)\, a_{l'm'}(\eta_0)}
&=  \frac{4\pi }{9}\,\delta_{ll'}\,\delta_{mm'}\,
 \int_{k_*(l)}^\infty 
 \frac{dk}{k}\,\Bigl(j_l(k\Delta\eta_*)\Bigr)^2\,P_\Psi (k)\notag\\
&\equiv  C_l\, \delta_{ll'}\,\delta_{mm'}.
\end{align}
When the spectral index $n(k)$ is constant, i.e., 
$P_\Psi (k) = P_0 \, k^{n-1}$ ($P_0$: constant), 
a simple calculation gives 
\begin{align}
C_l &=C_l^{(0)}\left(1 - \beta_l\right),\label{3.res}
\end{align}
where $C^{(0)}_l$ represents the angular power spectrum 
of the commutative case: 
\begin{align}
C_l^{(0)} &\equiv \frac{\pi^{3/2}}{9} (\Delta\eta_*)^{1-n} \,P_0\,
 \frac{\Gamma\left(\frac{3-n}{2}\right) 
 \Gamma\left( l + \frac{n-1}{2}\right) }
{\Gamma\left( \frac{4-n}{2}\right) \Gamma\left( l + \frac{5-n}{2}\right) },
\end{align}
and the deviation is expressed by
\begin{align}
 \beta_l &\equiv 
  \frac{4}{\sqrt{\pi}} 
  \frac{\Gamma\left(\frac{4-n}{2}\right) 
  \Gamma\left(l + \frac{5-n}{2}\right)}{\Gamma\left( \frac{3-n}{2}\right) 
  \Gamma\left(l+ \frac{n-1}{2}\right) }
  \int_0^{k_*(l)\,\Delta\eta_*} dx\, x^{n-2}\,\bigl(j_l(x)\bigr)^2. 
\end{align}
When $n=1$, $C_l^{(0)} = \text{const.}\bigl[l(l+1)\bigr]^{-1}$ 
and thus $l(l+1)C^{(0)}_l$ does not depend on $l$. 
However, 
the second term $\beta_l$ in \eqref{3.res} does depend on $l$,
and makes $l(l+1)C_l$ have a sharp damping at small $l$. 
The result with $n=0.95$ is depicted 
for various values of $l_s H$ in Fig.\ \ref{N=0.95}.

\resection{Conclusion and outlook}
In this paper, we considered an inflationary universe 
assuming that the geometry is expressed by a noncommutative space-time 
in the very early universe. 
We only take into account the noncommutativity of the angular coordinates 
$\Omega=(\theta, \varphi)$ 
since these should yield the most relevant effects 
on the deviation of the angular power spectrum from the standard value.
Instead of imposing boundary conditions, 
we solved the EOM by simply 
discarding those modes that are not allowed to exist in the subhorizon.

We calculated the two-point function. As depicted in Fig.\ \ref{N=0.95},
the result shows that the angular power spectrum certainly 
has a damping for small $l$. 
This has the same features with those observed in the WMAP and the COBE. 
This damping is usually interpreted based on the cosmic variance, 
but we showed that it has a possibility to be explained 
as effects of the noncommutativity of space-time during the inflationary era.

In this paper we have considered only scalar perturbations, 
but the above derivation can be used for any Gaussian fluctuations 
in the exponentially expanding universe, 
so that the tensor fluctuations will also have 
the same shape of damping at small $l$.
We expect that this behavior will be observed in experiments 
of the near future.

As stated above, instead of solving the EOM with proper boundary conditions  
we made an analysis of the classical solution in the superhorizon simply 
by discarding the modes 
that do not exist in the subhorizon. 
It should be enough for a qualitative argument, 
but in order to obtain a precise prediction that can be compared to 
the experimental data, 
we need to solve the problem by carefully choosing boundary conditions. 
Work in this direction is now in progress 
and will be reported in the subsequent paper \cite{fkm2}.

\section*{\large{Acknowledgment}}

The authors would like to thank T.~Higaki, W.~Hikida, H.~Kawai, 
N.~Maekawa, M.~Ninomiya and T.~Yamashita for discussions.  
This work is supported in part by Grant-in-Aid (No.\ 15540269) 
from the Japan Ministry of Education, Culture, Sports, 
Science and Technology, 
and by Grant-in-Aid for the 21st Century COE 
``Center for Diversity and Universality in Physics."


\baselineskip=0.7\normalbaselineskip


\begin{thebibliography}{99}


\bibitem{Bennett:2003bz}
C.~L.~Bennett {\it et al.},
arXiv:astro-ph/0302207.

\bibitem{Spergel:2003cb}
D.~N.~Spergel {\it et al.},
arXiv:astro-ph/0302209.

\bibitem{Peiris:2003ff}
H.~V.~Peiris {\it et al.},
arXiv:astro-ph/0302225.


\bibitem{cobe}
J.~R.~Bond, A.~H.~Jaffe and L.~Knox,
Phys.\ Rev.\ D {\bf 57} (1998) 2117
[arXiv:astro-ph/9708203].


\bibitem{Guth:1981}
A.~H.~Guth,
Phys. Rev. D {\bf 23} (1981) 347.

\bibitem{Sato1981}
K.~Sato,
Phys.~Lett.\ B {\bf 99} (1981) 66; 
Mon.~Not.~R.~Astron.~Soc.\ {\bf 195} (1981) 467.


\bibitem{Gratton:2003pe}
S.~Gratton, J.~Khoury, P.~J.~Steinhardt and N.~Turok,
arXiv:astro-ph/0301395.



\bibitem{Khoury:2003vb}
J.~Khoury, P.~J.~Steinhardt and N.~Turok,
arXiv:astro-ph/0302012.


\bibitem{Fukuma:2003kv}
M.~Fukuma, H.~Kawai and M.~Ninomiya,
arXiv:hep-th/0307061.



\bibitem{Yokoyama:1998rw}
J.~Yokoyama,
Phys.\ Rev.\ D {\bf 59} (1999) 107303.


\bibitem{Kawasaki:2003dd}
M.~Kawasaki and F.~Takahashi,
arXiv:hep-ph/0305319.

\bibitem{Chu:2000ww}
C.~S.~Chu, B.~R.~Greene and G.~Shiu,
Mod.\ Phys.\ Lett.\ A {\bf 16} (2001) 2231
[arXiv:hep-th/0011241];
R.~Easther, B.~R.~Greene, W.~H.~Kinney and G.~Shiu,
Phys.\ Rev.\ D {\bf 67} (2003) 063508
[arXiv:hep-th/0110226];
S.~Shankaranarayanan,
Class.\ Quant.\ Grav.\  {\bf 20} (2003) 75
[arXiv:gr-qc/0203060];
F.~Lizzi, G.~Mangano, G.~Miele and M.~Peloso,
JHEP {\bf 0206} (2002) 049
[arXiv:hep-th/0203099];
G.~L.~Alberghi, R.~Casadio and A.~Tronconi,
Phys.\ Lett.\ B {\bf 579} (2004) 1
[arXiv:gr-qc/0303035];
Q.~G.~Huang and M.~Li,
JHEP {\bf 0306} (2003) 014
[arXiv:hep-th/0304203];
J.~Martin and R.~Brandenberger,
Phys.\ Rev.\ D {\bf 68} (2003) 063513
[arXiv:hep-th/0305161];
Q.~G.~Huang and M.~Li,
arXiv:astro-ph/0311378.




\bibitem{Sachs}
R.~Sachs and A.~Wolfe,
Astrophys.\ J.\ {\bf 147} (1967) 73.


\bibitem{Brandenberger:2002nq}
R.~Brandenberger and P.~M.~Ho,
Phys.\ Rev.\ D {\bf 66} (2002) 023517
[AAPPS Bull.\  {\bf 12N1} (2002) 10]
[arXiv:hep-th/0203119].


\bibitem{Huang:2003zp}
Q.~G.~Huang and M.~Li,
JHEP {\bf 0306} (2003) 014
[arXiv:hep-th/0304203].


\bibitem{q}
S.~Cremonini,
arXiv:hep-th/0305244.


\bibitem{Tsujikawa:2003gh}
S.~Tsujikawa, R.~Maartens and R.~Brandenberger,
Phys.\ Lett.\ B {\bf 574} (2003) 141
[arXiv:astro-ph/0308169].


\bibitem{Huang:2003hw}
Q.~G.~Huang and M.~Li,
JCAP {\bf 0311} (2003) 001
[arXiv:astro-ph/0308458].


\bibitem{fkm2}
M. Fukuma, Y. Kono and A. Miwa,
hep-th/0312298. 


\bibitem{hppe1982}
J.~Hoppe, PhD Thesis (1982), Soryushiron Kenkyu (Kyoto) {\bf 80} (1989) 145.

\bibitem{Hoppe:vv}
J.~Hoppe and H.~Nicolai,
Phys.\ Lett.\ B {\bf 196} (1987) 451.

\bibitem{deWit:1988ig}
B.~de Wit, J.~Hoppe and H.~Nicolai,
Nucl.\ Phys.\ B {\bf 305} (1988) 545.


\bibitem{Madore}
J.~Madore,
Class. Quantum Grav. {\bf 9} (1992) 69.


\bibitem{Liddle}
A.~R.~Liddle, D.~H.~Lyth,
``Cosmological Inflation and Large-Scale Structure,'' \\
Cambridge University Press.
 


\bibitem{Bardeen:kt}
J.~M.~Bardeen,
Phys.\ Rev.\ D {\bf 22} (1980) 1882.


\bibitem{Kodama:bj}
H.~Kodama and M.~Sasaki,
Prog.\ Theor.\ Phys.\ Suppl.\  {\bf 78} (1984) 1.

\bibitem{Brandenberger:1992}
V.~F.~Mukhanov, ~H.~A.~Feldman ~and ~R.~H.~Brandenberger,\\
Phys.\ Rept.\ {\bf 215}\ (1992) 203. 




\end{thebibliography}
\end{document}